\documentclass[conference]{IEEEtran}
\IEEEoverridecommandlockouts
\usepackage{cite}
\usepackage{amsmath,amssymb,amsfonts}
\usepackage{algorithmic}
\usepackage{graphicx}
\usepackage{textcomp}
\usepackage{xcolor}
\usepackage{comment}
\usepackage{tabularx}
\usepackage{booktabs}
\usepackage{multirow}
\usepackage{balance}
\usepackage{array}
\newcolumntype{C}[1]{>{\centering\arraybackslash}p{#1}}

\def\BibTeX{{\rm B\kern-.05em{\sc i\kern-.025em b}\kern-.08em
    T\kern-.1667em\lower.7ex\hbox{E}\kern-.125emX}}
    
\begin{document}

\title{Designing AI Tutors for Interest-Based Learning: Insights from Human Instructors}

\author{\IEEEauthorblockN{Abhishek Kulkarni\IEEEauthorrefmark{1},
Sharon Lynn Chu\IEEEauthorrefmark{2}}
\IEEEauthorblockA{Embodied Learning and Experience Lab,
University of Florida\\
Gainesville, FL USA\\
Email: \IEEEauthorrefmark{1}kulkarniabhishek@ufl.edu,
\IEEEauthorrefmark{2}slchu@ufl.edu}}

\maketitle

\begin{abstract}
Interest-based learning (IBL) is a paradigm of instruction in which educational content is contextualized using learners' interests to enhance content relevance.
IBL has been shown to result in improved learning outcomes.
Unfortunately, high effort is needed for instructors to design and deliver IBL content for individual students.
LLMs in the form of AI tutors may allow for IBL to scale across many students.
Designing an AI tutor for IBL, however, first requires an understanding of how IBL is implemented in teaching scenarios.
This paper presents a study that seeks to derive this understanding from an analysis of how human instructors design and deliver IBL content. 
We studied 14 one-to-one online tutoring sessions (28 participants) in which tutors designed and delivered a lesson tailored to a student’s self-identified interest.
Using lesson artifacts, tutoring transcripts, interviews, and questionnaires, findings include themes on how tutors integrate interests during instruction and why. 
Finally, actionable design implications are presented for LLM-powered AI tutors that aim to deliver IBL at scale.
\end{abstract}

\begin{IEEEkeywords}
interest-based learning, AI tutor, large language models, pedagogy-informed design, qualitative analysis
\end{IEEEkeywords}

\section{Introduction}
Interest-based learning (IBL) has gained attention as an educational approach for its potential to impact learning outcomes such as motivation, retention, and self-efficacy positively \cite{kulkarni2022interest, aguar2017sail}.
In essence, IBL seeks to contextualize educational content around learners' interests \cite{hidi2006four}.
For instance, a tutor teaching Newton’s Laws of Motion to a student passionate about skateboarding may ask, \textit{``why do you need to push off the ground to start moving when you skateboard?''} to introduce Newton’s First Law. 
The student can then discuss momentum and friction by analyzing how different surfaces affect a skateboard’s speed, connecting physics concepts to something that they are inherently interested in.
However, in practice significant effort is needed for instructors to design such personalized content \cite{kayalar2017study}.
This has limited a wider adoption of IBL.
An instructor may be able to gauge one student's interest and use it to personalize their lessons, but this increasingly becomes untenable as the number of students grows.

Large language models (LLMs) present a path to scaling IBL where LLM-powered AI tutors could generate and deliver individualized interest-based explanations in real time \cite{graefen2024chat}.
However, tuning LLMs to create AI tutors for IBL needs careful design.
For example, prompts given to LLMs, if not properly designed, can yield shallow engagement or incoherent instruction \cite{rodriguez2024influence}.
We propose that an in-depth understanding of how human instructors design and deliver IBL can provide insights that can be used for the careful design of LLM-powered AI tutors for IBL.
Thus the study presented in this paper examines how educators, when explicitly prompted to do so, integrate students' interests and what they use interests for when designing and teaching IBL content in one-to-one online tutoring contexts.
Our work contributes an understanding of how human instructors implement IBL in the form of functions and mechanisms of interest integration in teaching, and concrete implications for the design of LLM-powered AI tutors for IBL.

\section{Background and Related Work}
\subsection{IBL and the Challenge of Scale}
IBL contextualizes educational content using learners’ interests \cite{hidi2006four}. 
The four-phase model of interest development \cite{hidi2006four} describes four types of interest: triggered situational interest, maintained situational interest, emerging individual interest, and well-developed individual interest.
On one end, triggered situational interest is elicited using external factors in a learning environment, while on the other end, well-developed individual interest is something an individual has engaged with over a long period of time.
The focus of our work is on well-developed individual interest as they already hold intrinsic value for the learner, reducing the need for constant external reinforcements.
This kind of interest is oftentimes conceptualized as one's hobbies \cite{edelson2004interest}.

Many studies across online and in-person contexts have shown that IBL can result in improved comprehension, retention, motivation, engagement, and self-efficacy (e.g., \cite{aguar2017sail, kulkarni2022interest, beh2018achieving}).
For example, SAIL \cite{aguar2017sail} is an online learning system 
that contextualized computer science exercises using pre-defined interest categories (e.g., sports, entertainment) and manually authored multiple versions of each item.
The semester-long study (307 learners) conducted using the system reported reduced gender performance disparities and high perceived understanding.
More broadly, most IBL interventions have relied on manually created, educator-driven materials.
For instance, interests have been embedded into math word problems \cite{rembert2019exploring} and technology to support older adults \cite{beh2018achieving}. 

Such studies integrate interests in bounded components such as quizzes rather than lesson-level instruction, and make use of categorized student interests.
With these constraints, scaling IBL to varied topics across an academic year would involve eliciting interests, generating tailored content, and maintaining instructional coherence, making it hard to sustain \cite{kayalar2017study}. 

\subsection{Intelligent Tutoring Systems (ITS) and LLMs}
Work in ITS suggests that well-structured online tutoring can yield engagement and learning comparable to in-person instruction \cite{chu2018effects}. 
In prior ITS work, personalized instruction has been achieved via difficulty adjustment, adaptive hints, and misconception handling \cite{graesser2005autotutor}.
A few systems such as SAIL \cite{aguar2017sail} have looked at integrating instruction personalization by using students' interests, but instruction by these systems are domain-specific, largely limited to problems or quizzes, and tend to rely on scripted dialogue.
LLMs offer a path to scaling IBL through their ability to generate adaptive, context-aware content in almost real-time and delivering it through dialogue \cite{park2024empowering}. 
LLMs may extend ITS beyond single-domain tutoring, but they still struggle with engagement, pedagogical coherence, and mismatched interaction expectations \cite{rodriguez2024influence}.
Thus, designing LLMs for effective tutoring requires a solid understanding of how the desired tutoring is performed.

\section{Our Approach}
A longstanding approach in ITS research is to model human tutoring strategies to improve adaptive instruction and feedback.
For example, AutoTutor was developed by analyzing naturalistic human tutoring to extract recurring moves (like feedback or hints) and then implementing those moves in a rule-based dialogue manager; across studies, AutoTutor has shown substantial learning gains (reported around 0.8 standard deviations over reading-only controls) \cite{graesser2005autotutor}.
Another example is CIRCSIM-tutor, a dialogue-based ITS for cardiovascular physiology that models human tutors by extracting pedagogical and discourse-planning rules.
In evaluations, it produced larger learning gains than reading text, and later comparisons found performance on par with human tutoring while out-performing a no-tutoring condition \cite{woo2006intelligent}.

Similarly for IBL then, we propose that the study of how human tutors engage in instruction through the use of students' interests is needed to inform the design of LLM-based AI tutors.
While not implemented on a large scale, IBL is a technique that instructors naturally resort to in their practical teaching on an \textit{ad hoc} basis. 
For instance, classroom observations show that teachers frequently incorporate relevance information, especially references to students' experiences, without being instructed to do so \cite{chu2023relevant}.
Teachers also draw on students' ``funds of identity'' \cite{esteban2014funds}, which are practices meaningful for students for making instruction relevant.
Similar efforts can be seen from teachers using students' out-of-school interests within classroom tasks \cite{tadic2019my}.

If explicitly prompted to do so, IBL then entails an instructional approach that instructors are typically able to implement independently.
In our approach, we analyze how human tutors use given student interests to teach given topics, and identify the most frequently used strategies across all tutors to derive implications for the design of IBL AI tutors.

\section{Research Questions and Study Description}
Given a topic to teach and a student's interest,

\noindent \textbf{RQ1}: \textit{What do tutors use student interests for in instruction?}

\noindent \textbf{RQ2:} \textit{How do tutors integrate student interests in instruction?}

\noindent \textbf{RQ3:} \textit{How do students and tutors perceive the integration of student interests in instruction?}

\subsection{Participants}
Twenty-eight participants (14 unique tutor-student pairs) were recruited via flyers and a university participant pool.
The sample size is consistent with prior qualitative studies that analyze tutoring interactions \cite{seo2008conversation, walker2022tackling}.
Each session lasted ~90 minutes for tutors (7 female, 7 male; mean age: 26 years) and 45 minutes for students (7 female, 7 male; mean age: 23 years).
To assess teaching-related comfort for tutors, we administered the Teaching Anxiety Scale (TCHAS), finding an average score of 33/70, indicating low anxiety ($\leq$ 45, range: 21-40) \cite{parsons1973assessment}.
Five tutors had 6 months of teaching experience, 6 had between 12 and 36 months, and 3 had more than 36 months.
Tutors came from diverse fields, including engineering, computer science, sociology, and psychology.

\subsection{Study Procedure}
The study was approved by the university's ethics board and conducted on Zoom with randomly assigned tutor-student pairs.
% Signed consent was obtained before the start of the study.
Tutors were recruited first and then a student with matching availability was selected. 
One day prior to the study, students completed an emailed Qualtrics questionnaire capturing their interest in the form of hobbies \cite{edelson2004interest} and pre-existing knowledge on selected educational topics.

During the session, tutors joined first, signed consent, and completed a questionnaire (TCHAS; teaching experience). 
The tutor was then briefed that their task was to design and teach a lesson using a student's stated hobby.
They then selected a concept they felt confident teaching from a list derived from topic familiarity ratings (1-5) given by the student matched with them.
Concepts for which the matched student reported low familiarity (score $\leq$ 2) were shortlisted, and the tutor chose from these options (Maslow’s hierarchy of needs; basics of quantum mechanics; internet protocol; entropy and the second law of thermodynamics; and hippocampus and synaptic consolidation). 

After topic selection, the student's hobby was shared and tutors were given pre-collected reference material (from publicly available textbooks) to bound the concept scope.
Tutors were told to assume that the student had no prior knowledge and received no lesson structure guidelines beyond creating a plan to use as a teaching aid.
They had 50 minutes to design the lesson, and were allowed to conduct online research but not to use AI tools. 
Their screen was shared and recorded.
After designing, tutors participated in an interview reflecting on their approach. 

The student then joined and both participants were instructed to treat the interaction as a one-to-one tutoring session and to interact freely. 
Tutors delivered the designed IBL lesson, with the session recorded (avg = about 30 minutes).
After teaching, students moved to a breakout room to complete a questionnaire gauging their experience and understanding. 
They completed a three-question quiz (two Multiple-Choice Questions and one open-ended explanation).
Meanwhile, tutors participated in a post-teaching interview, were then debriefed, and exited the session.
Finally, students were interviewed about their learning experience and debriefed, marking the end of the study.

\section{Data Analysis}
Data sources included lesson plans, lesson design interviews, tutoring session transcripts, and post-session interviews with both tutors and students.
Qualitative coding was conducted in MAXQDA where all coding was done manually.

We used inductive open coding.
One researcher coded lesson plans and all interviews line-by-line to capture lesson structure and content, tutors' rationale and intended interest use, tutors' reflections post-session, and students' perceptions of the lesson and interaction.
Tutoring session transcripts were coded utterance-by-utterance (each student or tutor turn) to capture instances of interest integration.

In a second cycle, two researchers conducted axial coding on the open codes and organized them into higher-order categories.
Any discrepancies were resolved through discussion.
We triangulated across data sources to ensure that intended interest integration was supported in the enacted interaction.
Tutor-reported strategies (post-design) were traced to where they appeared in lesson plans and how they were carried out in the transcripts, then compared against tutor/student reflections. 

This analysis produced three sets of findings aligned with our RQs.
First, functions (RQ1) captured why tutors invoked interests (derived primarily from tutoring transcripts, confirmed in tutor rationales and reflections). 
Second, mechanisms (RQ2) captured how interests were integrated (derived from the use of interest moves in lesson plans and transcripts).
And third, perceptions (RQ3) captured tutors' and students' reflections on the IBL teaching and learning experience.

For reliability, a second researcher independently coded 30\% of the tutoring transcripts only using the finalized codebook (after completing axial coding) yielding an inter-rater reliability calculated using Cohen’s kappa = 0.76, indicating substantial agreement.

\section{Findings}
\subsection{RQ1: Functions of Interest Integration} 
Tutors used student interests for three main functions:

\noindent \textbf{Function 1: To build rapport (12 of 14 sessions).} Tutors leveraged students' interests early in the session to establish a connection and create a comfortable learning environment, typically through follow-up questions that invited the student to elaborate (e.g., \textit{``Tell me more about why you prefer fantasy over other genres like nonfiction or sci-fi?''}).
Tutors also described this as intentional rapport work in post-study interviews, \textit{``when beginning to create the lesson plan, I started with the snippet of what they had written [referring to the interest]. I also felt it was important to give them a chance to talk about it more.''}

\noindent \textbf{Function 2: To Introduce Concepts (12 of 14 sessions).}
Tutors used interests as an entry point for the learning topic, treating the interest as part of the learner’s prior knowledge and using it to bridge into new material.
These moves were sometimes explicit, as when a tutor framed an abstract psychology idea through an art-focused student’s hobby, \textit{``that concept is a bit abstract, so let’s put it in terms of art, which you’re familiar with.''}
Other times the integration was more implicit, where the tutor began explaining a new concept directly inside the hobby context, \textit{``...imagine another similar triangle which has a hierarchy, which has like five layers and we call it the triangle of clicking good pictures.''}

\noindent \textbf{Function 3: To Hand Over the Dialogue to Students (9 of 14 sessions).}
Tutors used interests to shift turn-taking and invite student agency.
They posed interest-grounded prompts to encourage reflection, prompting students to articulate their understanding.
For example, while teaching entropy to a fiction-interested student, one tutor asked, \textit{``once gas particles escape the container and fill the room, there’s no reason for them to return just like the kingdom's peasants in our story. Why would they want to go back?''}
Such turn-transfer often happened after the core concept was explained and when tutors were looking to gauge understanding.

\subsection{RQ2: Mechanisms of Interest Integration} Tutors employed four mechanisms to integrate interest:

\noindent \textbf{Mechanism 1: Peripheral vs. Central Integration (11 central and 3 peripheral sessions).}
Two approaches to integrating student interests emerged.
In peripheral integration, tutors first developed the conceptual explanation and then inserted interest connections at specific points. 
One tutor summarized this as, \textit{``I had my content ready, and then I looked for points where I could bring in their interest.''} 
This approach was common when tutors were less familiar with the interest or when direct connections were difficult, leading to short student turns (e.g., \textit{``that makes sense''}). 
In central integration, the interest became the organizing frame for the lesson, e.g., a sports interest structuring the full explanation of Maslow’s hierarchy of needs.

\noindent \textbf{Mechanism 2: 
Contextualizing Scenarios (10 of 71 interest integration instances).
} 
Tutors created interest-grounded scenarios to prompt reasoning or reflection, often as open-ended questions (e.g., \textit{``...think of playing a guitar. What elements of entropy do you notice?''}). 
Designing scenarios was harder when the interest and topic had limited conceptual overlap. 
In such cases, tutors broadened the interest; reflecting on a Maslow’s lesson tied to painting, one tutor noted, \textit{``I’m not really experienced with painting, that’s why I had to make it more broad [broadening the context to art in general]''}.

\noindent \textbf{Mechanism 3: 
Illustrative Examples (23 of 71 interest integration instances).}
When topic-interest overlap was more direct, tutors used interest-based examples to clarify and reinforce concepts, often introduced mid-explanation to solidify understanding.
For instance, explaining synaptic consolidation to a student interested in gardening, a tutor said, \textit{``with gardening, you might start by looking up whether watering benefits a plant, but as you care for it, the process becomes second nature, and you no longer need to look things up.''} 
However, when tutors lacked knowledge of the interest, examples could feel forced and potentially distracting. 
As one tutor put it, \textit{``I was thinking about how to connect it in layman terms, like football. I’m not a sports guy, so I might not be able to come up with the best examples.''}

\noindent\textbf{Mechanism 4: 
Analogical Bridges (38 of 71 interest integration instances).}
Tutors relied on analogies when direct connections were tenuous, using the interest to build an analogical bridge to abstract concepts. 
One tutor teaching entropy used a fiction-based analogy, \textit{``...I'm currently in my lord of the rings era. And you know its just like this gravitation towards the idea that this small confined state is not enough. Like I want to expand, spread the chaos.''} 
Tutors noted that analogies required careful design to remain accurate while still intuitive, \textit{``...one challenge was ensuring the analogies didn’t feel forced or disconnected from the core concept.''} 
Despite this challenge, analogy-based integration was the most frequently used mechanism, largely as it enabled connection even when there was no obvious \textit{``organic''} overlap.

\begin{table*}[t]
\centering
\caption{Design implications for an interest-based AI tutor. 
\textbf{A}=\emph{Opening/rapport-building}, \textbf{B}=\emph{Instruction \& explanation}, \textbf{C}=\emph{Evaluation \& formative assessment}.}
\label{table:design-implications}
\renewcommand{\arraystretch}{1.05}
\setlength{\tabcolsep}{3pt}
\footnotesize
\begin{tabularx}{\textwidth}{@{}p{0.55cm}p{4.6cm}X@{}}
\toprule
\textbf{Phase} & \textbf{Themes} & \textbf{Design implications} \\ 
\midrule

\multicolumn{3}{@{}l}{\textbf{RQ1:} \textit{What did tutors use student interests for?}}\\
\multicolumn{1}{c}{A} & Build rapport
& \textbf{Prompt about Interest:} Prompt open-ended questions about the learner's interest at session start and ask follow-up questions to gain more contextual information. \\

\multicolumn{1}{c}{B} & Introduce concepts
& \textbf{Frame Concept with Interest:
} Frame topics using the student’s interest as a conceptual bridge early on in session. \\

\multicolumn{1}{c}{C} & Hand over dialogue to student
& \textbf{Craft Reflection Prompts using Interest:} Use interest-driven questioning to shift agency to the student and prompt reflection on newly introduced concepts. \\

\midrule

\multicolumn{3}{@{}l}{\textbf{RQ2:} \textit{How did tutors integrate student interests?}}\\
\multicolumn{1}{c}{B} & Central vs.\ peripheral integration
& \textbf{Adaptively Switch Depth of Interest Integration:} Default to central integration. Fall back on peripheral when alignment/context is weak. \\

\multicolumn{1}{c}{B} & Analogical bridges
& \textbf{Use Interest-based Analogies as Conceptual Bridges:} Create analogies when direct connections are infeasible. Ensure thereafter that misconceptions were not introduced. \\

\multicolumn{1}{c}{B} &  Illustrative examples
&  \textbf{Use Interest-based Examples for Reinforcing Learning:} Provide interest-based examples to reinforce and clarify taught concepts. \\

\multicolumn{1}{c}{C} & Contextualizing scenarios
& \textbf{Use Interest-based Scenarios for Active Reasoning:} Use interest in the form of reflective scenario-based questions to prompt active reasoning. \\

\bottomrule
\end{tabularx}
\end{table*}

\subsection{RQ3: Perceptions of Interest Integration}
Students described IBL as making the lesson feel \textit{``more casual''} and easier to approach because they were \textit{``familiar with the grounds''}. 
One student noted, \textit{``I like that the tutor used an example from one of the games that I’ve played,''} while another said, \textit{``as soon as she mentioned books, my mind went to a book that I had read which directly related to the topic.''} 
Students also linked interests directly to engagement, \textit{``I think the use of my hobby kept me interested in the topic,''} and \textit{``she gave me examples that related to my hobby and that made me happy already...it was like work and play at the same time.''} 

Tutors generally valued interest integration because it helped the session feel interactive, \textit{``not one-way communication...it felt like a conversation''.}
They used student responses as feedback, highlighting how students \textit{``built off my thoughts''} and how they could explain ideas in their own words.
Tutors also highlighted a challenge of keeping the interest-topic link coherent, especially when alignment was weak or the tutor was unfamiliar with the hobby. 
This showed up as difficult transitions to/from examples (\textit{``my transition wasn’t that smooth''}) or concerns that connections became \textit{``superficial, not as well developed''}. 
In hindsight, many tutors wanted to pause more and invite student-generated examples (\textit{``ask the student...to get [an] example of their own''}).
Several also described interest integration as co-constructed in real time where student responses led to new examples (\textit{``from his responses I got that idea''}), and one noted that creating these connections \textit{``on the go''} would be \textit{``way easier''} with an LLM.

\section{Discussion}
This study reveals how tutors use learner interests during instruction and how they integrate them in dialogue.
We translate our findings into design implications (refer Table \ref{table:design-implications}) by framing them within the three phases commonly described for tutoring dialogue structure \cite{rus2015automated} in prior work.
The three phases are: (A) opening/rapport-building; (B) instruction and explanation; and (C) evaluation such as assessment/metacognition.
We describe these implications below.

\textbf{Phase A. Opening/rapport-building.}
Tutors most consistently began by asking about the learner’s hobby to establish rapport and elicit richer participation.
This aligns with existing evidence that instructor-learner rapport supports engagement and learning \cite{frisby2010instructor}. 
Accordingly, an AI tutor should \textbf{prompt about interest} by asking open-ended questions about the learner's hobby at session start and using follow-up questions to gather contextual details that can be reused later in the session.

\textbf{Phase B. Instruction and explanation.}
Tutors also consistently used interests to introduce unfamiliar concepts, sometimes explicitly and sometimes by embedding explanations directly inside the hobby context.
This mirrors broader learning principles which show connecting new ideas to existing schemas can improve understanding and retention \cite{bransford2000people}.
Thus, an AI tutor should \textbf{frame concept with interest} by using the student’s hobby as a conceptual bridge early in the session, at the point of introducing the core idea.
At the lesson level, tutors varied in the \emph{depth} of integration.
Most sessions used central integration, suggesting an AI tutor should default to interest-as-an-organizing-frame when alignment and interest detail are available, while allowing peripheral integration when connections are tenuous or the interest is unfamiliar. 
When direct topic-interest overlap was limited, tutors relied most on analogical bridges and secondarily on illustrative examples.
Prior work notes analogies can make concepts accessible but also risk overextension and misconception \cite{glynn2012explaining}, implying to \textbf{use interest-based analogies as conceptual bridges} while verifying that misconceptions were not introduced. 
In contrast, examples functioned mainly as brief clarifiers, consistent with worked-example effects that can reduce cognitive load \cite{sweller2006worked}, suggesting \textbf{use interest-based examples for reinforcing learning} to reinforce and clarify taught concepts.

\textbf{Phase C: Evaluation and formative assessment.}
Beyond explanation, tutors used interests to shift agency to learners through interest-grounded reflection prompts that encouraged students to articulate and apply ideas.
This supported active learning, which has shown to promote understanding of concepts and result in students self-constructing knowledge rather than consuming it passively \cite{hartikainen2019concept}. 
Thus, the AI tutor should \textbf{craft reflection prompts using interest} to shift agency and prompt reflection on newly introduced topics.
Finally, scenarios were less frequent, often because they were harder to design under low knowledge of hobby; however, tutor reflections emphasized wanting more learner involvement (e.g., student-generated examples and activities). 
This suggests that \textbf{using interest-based scenarios for active reasoning} can serve as a targeted strategy for eliciting reasoning when implemented accurately, consistent with case-based learning and analogical transfer work showing that reasoning in familiar contexts can deepen understanding \cite{bassok2003analogical}.
Collectively, these implications describe a practical design space for interest-based AI tutors.

\section{Limitations, Future Work, and Conclusion} \label{sec:conclusion}
First, the study took place in a controlled setting, which may not reflect real-world instructional pacing.
Future work should examine interest integration in more naturalistic contexts. 
Second, our findings are specific to adult learners in one-to-one online tutoring. 
IBL may look different with younger learners who need more scaffolding, where attention and social dynamics differ.
Future research should investigate across age groups and learning environments to assess broader applicability.

Finally, while LLM-powered AI tutors may help scale IBL, they introduce ethical and practical risks, including privacy, inequitable access, bias in generated content, and over-reliance on AI. 
Future work should evaluate deployment strategies so interest-based AI tutors align with ethical and pedagogical best practices.
By translating this work's insights into design implications, our work lays the foundation for scalable IBL that preserves the depth and intentionality of human instruction.

%\section*{Acknowledgment}
\balance
\bibliography{bibliography.bib}
\bibliographystyle{IEEEtran}

\end{document}